\documentclass[%
 twocolumn,showpacs, amsmath,amssymb,fleqn,aps,prb,floatfix
]{revtex4}

\usepackage{graphicx}
\usepackage{dcolumn}
\usepackage{bm}
\usepackage{color}
\usepackage[mathlines]{lineno}
\usepackage{multirow}

\newcommand{\EF}{\ensuremath{E_{\rm F}}}
\newcommand{\vc}[1]{\ensuremath{\mathbf{#1}}}
\newcommand{\sqa}{\ensuremath{\hat{\vc{s}}}}

\begin{document}


\title{Spin-flip hot spots in ultrathin films of monovalent metals:
  Enhancement and anisotropy of the Elliott-Yafet parameter}

\author{Nguyen H. Long} 
\email{H.Nguyen@fz-juelich.de}
\affiliation{Peter Gr\"unberg Institut and Institute for Advanced Simulation, Forschungszentrum J\"ulich and JARA, D-52425 J\"ulich, Germany}
\author{Phivos Mavropoulos} \email{Ph.Mavropoulos@fz-juelich.de}
\affiliation{Peter Gr\"unberg Institut and Institute for Advanced Simulation, Forschungszentrum J\"ulich and JARA, D-52425 J\"ulich, Germany}
\author{Swantje Heers, Bernd Zimmermann, Yuriy Mokrousov} 
\affiliation{Peter Gr\"unberg Institut and Institute for Advanced Simulation, Forschungszentrum J\"ulich and JARA, D-52425 J\"ulich, Germany}
\author{Stefan Bl\"ugel} 
\affiliation{Peter Gr\"unberg Institut and Institute for Advanced Simulation, Forschungszentrum J\"ulich and JARA, D-52425 J\"ulich, Germany}

\begin{abstract}
  In contrast to the long-known fact that 
  spin-flip hot spots, i.e., special \vc{k}-points on the Fermi
  surface showing a high spin-mixing parameter, do not occur in the bulk
  of monovalent (noble and alkali) metals, we found them  on the
  surface Brillouin-zone boundary of ultrathin films of these metals.  Density-functional
  calculations within the Korringa-Kohn-Rostoker Green function method
  for ultrathin (001) oriented Cu, Ag, and Au films of 10-layer
    thickness show that the region around the hot spots can
  have a substantial contribution, e.g.\ 52\% in Au(001), to the
  integrated spin-mixing parameter, that could lead to a significant
  enhancement of the spin-relaxation rate or spin-Hall angle in thin films. Owing to
  the appearance of spin-flip hot-spots, a large anisotropy of the
  Elliott-Yafet parameter [50\% for Au(001)] is also found in these
  systems. The findings are important for spintronics applications in
 which noble-metals are frequently used and in which  the
 dimensionality of the sample is reduced. 
\end{abstract}

\maketitle


\section{Introduction}
The term {\it spin-flip hot spots} was coined by Fabian and Das
Sarma\cite{fabian98,fabian99} when, based on the Elliott-Yafet
mechanism,\cite{elliott54,yafet63} they predicted a high spin-flip
probability for electrons at certain special points on the Fermi
surface of bulk Aluminium. The hot spots  are  formed by the
spin-orbit interaction  in the vicinity of
band degeneracies or near-degeneracies, frequently occurring at the
Brillouin-zone boundaries, high-symmetry lines or points of accidental degeneracy. 
On the other hand, hot spots should be absent in monovalent
metals,\cite{fabian98,fabian99} because their Fermi surface is nearly
spherical and in most cases does not cross the Brillouin-zone edge.
 
These conclusions were derived for the
bulk.\cite{fabian98,fabian_jap99,fabian_jvst99} However, in the case
of ultrathin films spin-flip hot spots may occur even in monovalent
metals owing to the different shape of the surface Brillouin zone
compared to the bulk. This is the main conclusion of the present
paper, which we base on simple theoretical arguments and on a
verification for (001) oriented ultrathin crystalline films of Cu, Ag,
and Au as well as alkali metals. We adopt the structural model of
  free-standing films as generic for films that are deposited on, or
  sandwitched between, insulating materials, in particular concerning
  the electronic structure of the quantum-well states in the film.
Our main focus is on the noble metals because of their frequent
  usage in spintronics applications as wires, contacts or
  probes.
We investigate the contribution of the hot spots to the Elliott-Yafet
parameter\cite{elliott54,yafet63} $b^2$ (defined below) as well as to
its anisotropy.  Our findings should be accounted for in spintronics
applications where the hot spots play a role, such as for giant
magnetoresistance, spin-Hall effect and spin
dynamics.\cite{zutic04,spinhall,steiauf09} 

\section{Theory}
We proceed with a short summary of the theoretical
background.\cite{fabian98,elliott54,yafet63} In
non-magnetic systems with space-inversion symmetry and in the presence of
the spin-orbit coupling the Bloch states at any \vc{k}-point are at
least twofold degenerate and comprise a superposition of spin-up and
spin-down states that is frequently called spin mixing:
\begin{equation}
\begin{array}{l l}
\Psi^{+}_{\mathbf{k}}(\mathbf{r})=&\left[a_{\mathbf{k}}(\mathbf{r})\left|\uparrow\right>+b_{\mathbf{k}}(\mathbf{r})\left|\downarrow\right>\right]e^{i\mathbf{kr}}\\ \\
\Psi^{-}_{\mathbf{k}}(\mathbf{r})=&\left[a^*_{\mathbf{-k}}(\mathbf{r})\left|\downarrow\right>-b^*_{\mathbf{-k}}(\mathbf{r})\left|\uparrow\right>\right]e^{i\mathbf{kr}}.
\end{array}
\label{mixwavefunction}
\end{equation}
Here, the spinors $\left|\uparrow\right>$ and
$\left|\downarrow\right>$ are eigenvectors of the $z$-component Pauli
matrix $\sigma_z$ (given a chosen $z$-axis) and
$a_{\mathbf{k}}(\mathbf{r})$ and $b_{\mathbf{k}}(\mathbf{r})$ are the
lattice-periodic parts of the Bloch states and are denoted as the
  large and small spin-component, respectively, because they are
  chosen such that the norm of $a_{\mathbf{k}}$ is maximal and that of
  $b_{\mathbf{k}}$ is minimal (see below).  The spin-expectation
values of these partner states are
$\vc{S}(\vc{k})=\left<\Psi^{+}_{\mathbf{k}}|\frac{\hbar}{2}\boldsymbol{\sigma}|\Psi^{+}_{\mathbf{k}}\right>
=
-\left<\Psi^{-}_{\mathbf{k}}|\frac{\hbar}{2}\boldsymbol{\sigma}|\Psi^{-}_{\mathbf{k}}\right>$
with $\boldsymbol{\sigma}$ denoting the vector of Pauli matrices.
The index ``$+$'' refers to the state with maximal $z$-component of
the spin-expectation value $S_z(\vc{k})=\hbar (\frac{1}{2}-b_{\mathbf
  k}^2)$, where $b_{\vc{k}}^2:=\int d^3r|b_{\vc{k}}(\vc{r})|^2\in
[0,\frac{1}{2}]$ defines the space-integrated spin-mixing parameter.
Maximizing the value of $S_z(\vc{k})$ is done with respect to all
possible linear combinations of the two degenerate states at $\vc{k}$,
and it is obvious that one can choose a different pair of states in
Eq.~(\ref{mixwavefunction}) by maximizing the projection of
$\vc{S}(\vc{k})$ along any spin-quantization axis \sqa\ prescribed by
the experimental conditions (e.g.\ by an external magnetic field or by
the polarization direction of an injected spin
current).\cite{Zimmermann12,Mokrousov13} The procedure for finding the
wavefunctions that yield the maximal $\vc{S}(\vc{k})$ is given in the
appendix. We will return to this freedom of choice of \sqa\ below when defining
the anisotropy of the Elliott-Yafet parameter.  The relation
between the large and small 
components of $\Psi^{-}_{\vc{k}}$ and $\Psi^{+}_{\vc{-k}}$ implied in
Eq.~(\ref{mixwavefunction}), i.e. presence of the coefficients $a^*_{-\vc{k}}(\vc{r})$ and $-b^*_{-\vc{k}}(\vc{r})$ in the expression for $\Psi^{-}_{+\vc{k}}$, follows from time-reversal and
space-inversion symmetry.\cite{elliott54,yafet63} It is also
convenient to define the Elliott-Yafet parameter $b^2$ as the
Fermi-surface average:
\begin{equation}
b^2=\left<b_{\mathbf{k}}^2\right>_{\rm FS}=\frac{1}{n\left(\EF\right)}\frac{1}{V_{\rm BZ}}\int_{\rm FS}\frac{d\Omega_{\mathbf{k}}}{\hbar
  v_\mathbf{k}} b_{\mathbf{k}}^2,
\label{bsq}
\end{equation}
where $d\Omega_{\mathbf{k}}$ is the Fermi-surface element, $v_{\mathbf{k}}$
is the modulus of the Fermi velocity, $n(\EF)$ is the density of states at the Fermi
level \EF, and $V_{\rm BZ}$ is the Brillouin-zone volume (or area
in two dimensions).

The spin-mixing parameter $b_{\vc{k}}^2$ is the main quantity of
interest in the analysis of many spin-flip related phenomena, as it
reflects the deviation of a Bloch state from being a spin eigenstate.
When $b_{\vc{k}}^2$ happens to be large (close to $\frac{1}{2}$), then
the particular state has an almost completely mixed spin
character.\cite{fabian98,fabian99,fabian_jap99} Concerning spin
relaxation, an electron scattered into this state practically loses 
its spin character---for example, according to Elliott's
approximation,\cite{elliott54} the spin relaxation time $T_1$ is
inversely proportional to the Elliott-Yafet parameter
$b^2$. Concerning the spin-Hall effect, such a state with large
$b_{\vc{k}}^2$  is associated with a high value of the Berry curvature\cite{spinhall,gradhand11} 
and exhibits thus  a strong contribution to the  conductivity
tensor. 
The mixing $b_{\vc{k}}^2$ becomes large at, and close to, special points in the BZ, 
the spin-flip hot spots, that we are examining in the
present work.

We should point out that the values of 
$b^2_{\vc{k}}$, as well as the integrated $b^2$, depend on the
direction of the spin-quantization axis \sqa, because the matrix
elements of the spin-flip part of the spin-orbit operator between
Bloch states change with respect to the \sqa\ axis along which
$\vc{S}(\vc{k})$ is maximized.\cite{Zimmermann12,Mokrousov13,Long13} 
Therefore we define the anisotropy of the
Elliott-Yafet parameter as\cite{Zimmermann12}
\begin{equation}
\mathcal{A}[b^2] = \frac{\max_{\sqa}b^2(\sqa) - \min_{\sqa}b^2(\sqa)}{\min_{\sqa}b^2(\sqa)}
\label{eq:anisotropy}
\end{equation}
by considering the maximum and minimum value with respect to all
  directions of \sqa\ in the unit sphere.

For bulk materials, this anisotropy effect, its microscopic origin and its relation to hot spots was analyzed in Refs.~\onlinecite{Zimmermann12,Mokrousov13}.
It is obvious that the thin film geometry  breaks the cubic symmetry and two surfaces, or generally, interfaces appear, and we expect
that the value of $b^2$ will be different for spin-quantization axes \sqa\ chosen  in the film plane  ([100]-direction) or perpendicular to the plane
[001]. This has been shown explicitly and attributed mainly to surface
states for the transition-metal W(001) films in Ref.~\onlinecite{Long13}. 
Physical consequences of the anisotropy are, e.g., a variation of the spin-relaxation time or of
the spin-Hall conductivity with respect to the direction of
polarization of the spin current in the material, that corresponds to the
polarization direction  \sqa\ in the present theory.

The fact that spin-flip hot spots occur at points of
degeneracy or near-degeneracy, in particular
at the back-folded energy bands at the Brillouin-zone edge, follows
from a consideration of transitions from a band at energy $E_{\vc{k}}$
to a band at $E_{\vc{k}}+\Delta_{\vc{k}}$ of small inter-band
separation $\Delta_{\vc{k}}$ under the action of the spin-orbit
Hamiltonian.\cite{fabian98,fabian99,fabian_jap99} From these arguments
it also follows that monovalent metals should not show spin-flip hot
spots,\cite{fabian98,fabian99,fabian_jap99} since their almost spherical Fermi surface either does not
cross the Brillouin-zone boundary (as for all alkali metals
except Cs) or, if it crosses the boundary (as for the noble
metals, forming a neck around the $L$-point), then it does so in a way
that there occur no hot spots. The validity of the latter statement is shown and discussed in detail in Ref.~\onlinecite{fabian_jap99}.

\begin{figure}
\includegraphics[width=8cm]{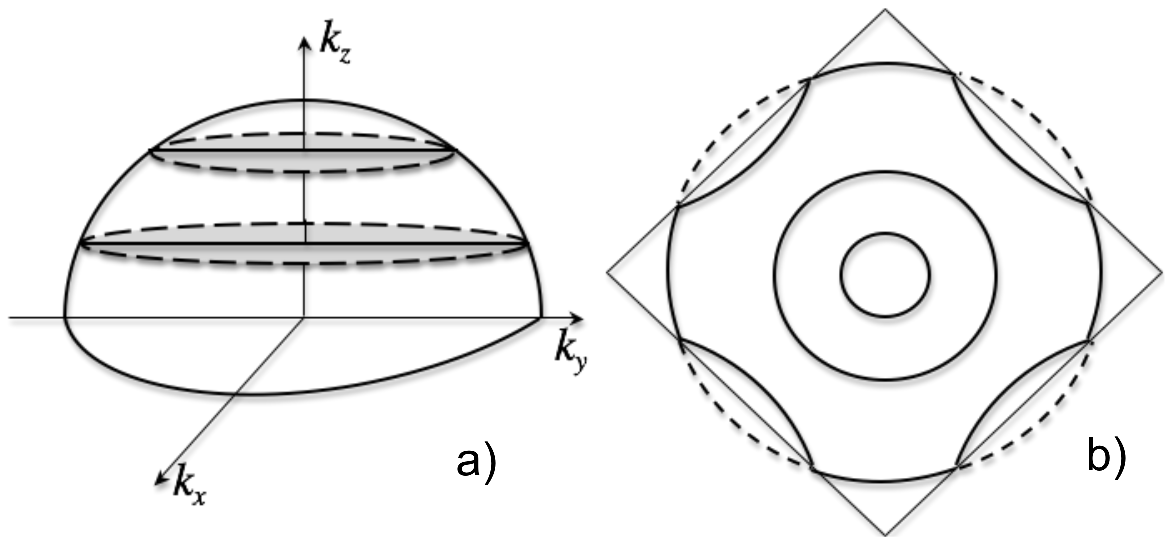}
\caption{Sketch of the ideal Fermi surface of a monovalent metal. (a)
  Spherical bulk Fermi surface; the circles shown at certain values of $k_z$
  correspond to the size-quantization of  $k_z$ in a film grown in the
  $z$-direction. (b) Fermi circles in a (001) film and the back-folding
  into the first surface Brillouin-zone, shown together with the crossed
  surface Brillouin-zone boundary.
  This schematic form of the Fermi surface is modified even in free-electron
  metals, especially if there is some {\it d} contribution at ${\mathrm E}_F$,
  as we see in Fig.~\ref{anibsqAu}.}
\label{fermisurface}
\end{figure}

However, the situation of monovalent metals changes when one considers
ultrathin films. 
In this case the two-dimensional periodicity implies
a surface Brillouin-zone, while the Fermi surface consists of rings
occurring when the spherical Fermi surface of the bulk system is cut
parallel to the surface plane at positions determined by the finite-size quantization of $k_z$ due to films of finite thickness. 
This well-known effect is schematically demonstrated in
Fig.~\ref{fermisurface}a. Considering a (001) film, where the surface
Brillouin-zone is a square, the larger of the Fermi rings exits the first
Brillouin zone and a back-folding occurs (see the dashed lines and
their back-folded counterparts in Fig.~\ref{fermisurface}b). 
The back-folding, under action of the periodic potential, can form a weak degeneracy lifting at the Brillouin-zone edge, with the resulting states being
energetically very close. But this is precisely the case when
spin-flip hot spots occur under the action of the spin-orbit coupling
between the near-degenerate states.

To what extent this actually happens in a realistic case depends on
the exact shape of the Fermi surface, the number of film layers, the
surface orientation, and of course the material. Here, we present {\it
  ab initio} results for Cu, Ag and Au (001) films of 10 layers
thickness, where we find always an effect of considerable magnitude.
This is in difference to 
alkali (001) thin films,  where we find that the magnitude of the effect depends on the film thickness due to the weak spin-orbit coupling.

The electronic structure is calculated within the local density
approximation to density-functional theory in the parametrization
  of Vosko {\it et al}.\cite{vosko} at experimental 
lattice parameters and ignoring surface relaxation. For our
calculations we employ the full-potential Korringa-Kohn-Rostoker Green
function method with an angular momentum cutoff of $l_{\rm
  max}=3$.  
The Fermi surfaces are interpolated from a mesh of $80\times 80$ points in the surface Brillouin zone, resulting in approximately 9000 \vc{k}-points on the Fermi surface.
Details on the formalism and implementation can be found in Refs.~\onlinecite{heers11,Papanikolaou02,Stefanou90}.

\section{Results and discussion}
Starting our analysis with the bulk properties, the spin-mixing
parameter for bulk Cu and Au is well studied (for $\sqa\parallel[001]$), e.g.\ in 
Ref.~\onlinecite{heers11} and Ref.~\onlinecite{gradhand09}. It is
long-known that the Fermi surfaces of the noble metals are
similar. The texture of $b_{\vc{k}}^2$ on the Fermi sufaces is also
similar, but the magnitude is very different due to the much stronger
spin-orbit coupling in Au. It is found that, for
  $\sqa\parallel[001]$, the value of $b^2_{\vc{k}}$ varies as a function of the position on the Fermi surface between 0.0005 and 0.002 in bulk Cu, 
between 0.0008 and 0.0025 in Ag, while in Au it clearly reaches much higher
values varying between 0.01 and 0.045. It is also found that there are
no hot spots on the noble-metal Fermi surfaces.  The Fermi-surface
average is calculated to be $b^2({\rm Cu})=0.0015$, $b^2({\rm
  Ag})=0.0017$, and $b^2({\rm Au})=0.03$.
We also find that in the bulk of noble metals
 the cubic symmetry and the absence of hot spots
makes the anisotropy of $b^2$ negligible (less than 0.1\% comparing
$\sqa\parallel$[001], [110] and [111]) but in the films it takes
large  values due to the hot spots, as we show below.

\begin{figure}
\includegraphics[width=8cm]{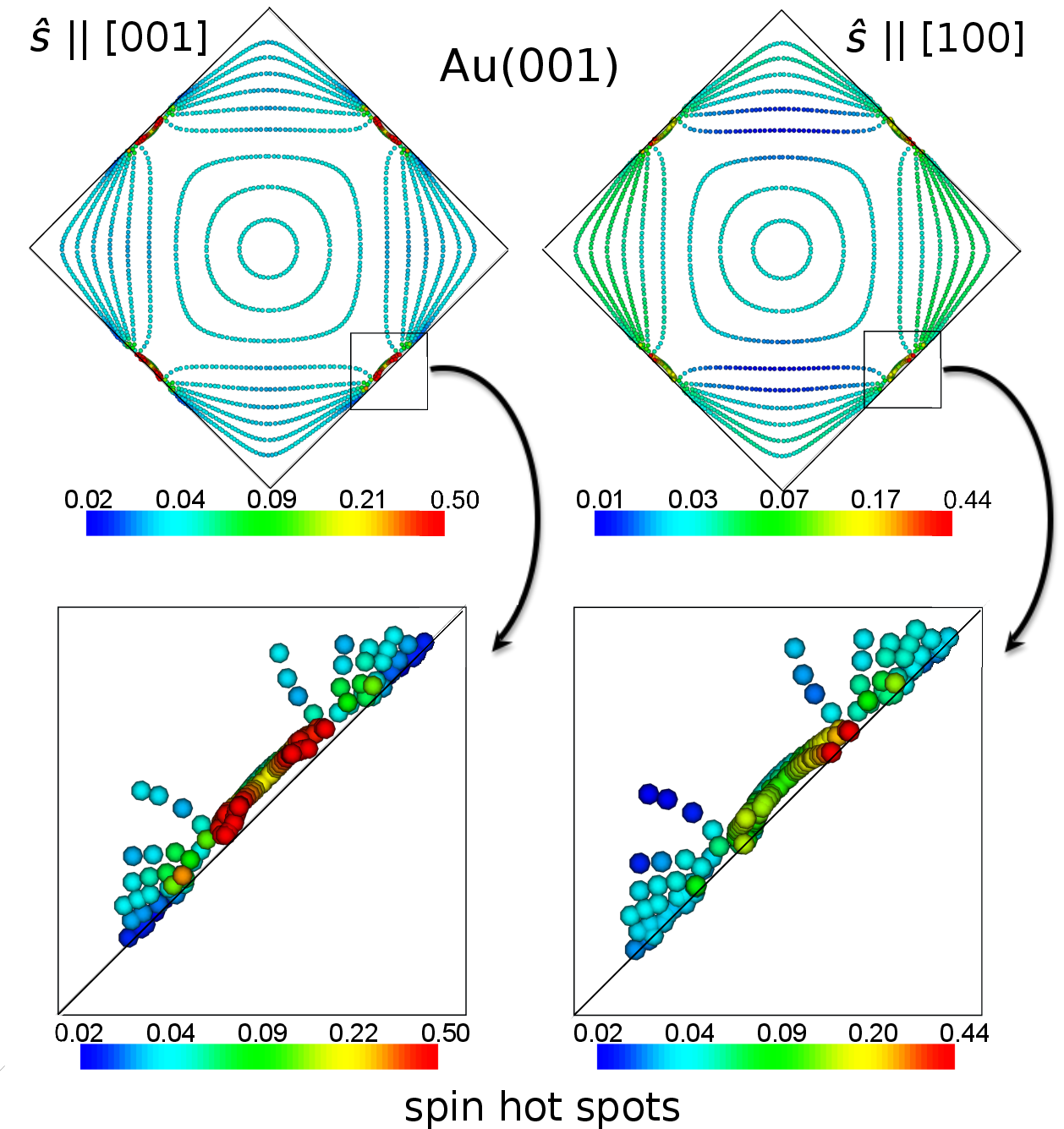} 
\includegraphics[width=8cm]{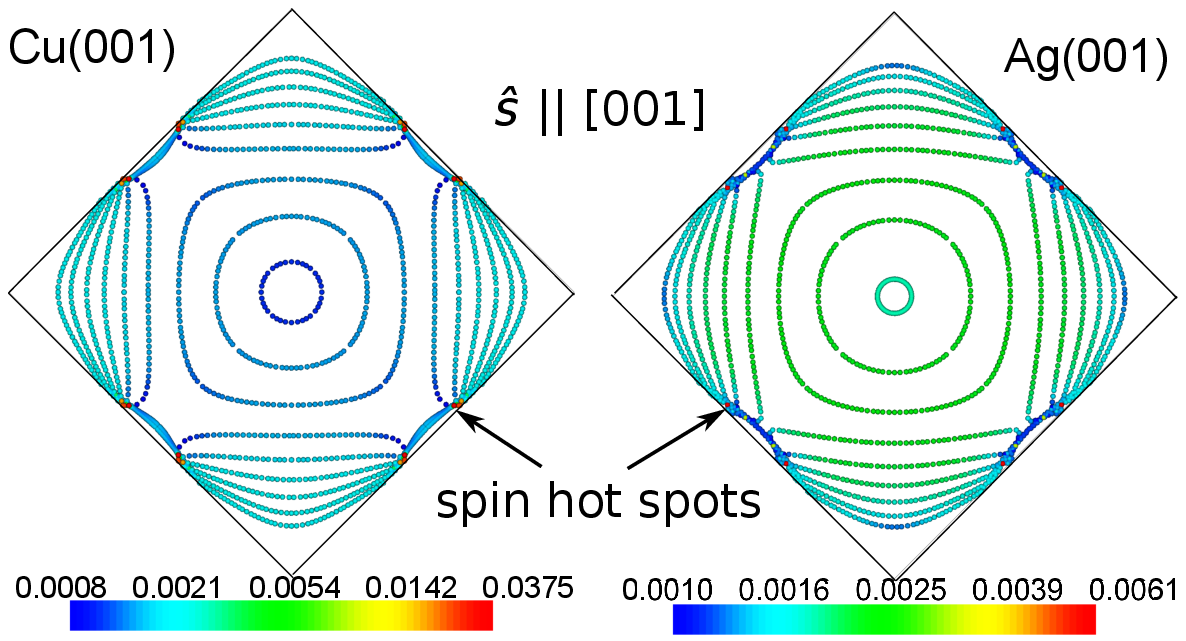} 
\caption{(Color online) Top: Distribution of spin-mixing parameter on
  the Fermi surface of a 10 atomic layer thick Au(001) film with the spin-quantization 
  axis along [001], i.e.\ out of plane (left panel) and
  in-plane along the [100] direction (right panel). The full surface
  Brillouin-zone is shown. Middle: Focus on the Brillouin-zone edge of
  Au(001) shown in the top panels in order to distinguish the extent
  of the hot spots. An asymmetry of distribution of $b_{\vc{k}}^2$
  can be seen in the case of $\sqa \parallel [100]$ (right). Bottom:
  Distribution of spin-mixing parameter on the Fermi surface of a
  Cu(001) (left) and a Ag(001) (right) 10-layer film with the
  $\sqa \parallel [001]$ (out of plane).}
\label{anibsqAu}
\end{figure}  

Now we examine the distribution of $b^2_{\vc{k}}$ on the Fermi
surfaces of the ultrathin films. Fig.~\ref{anibsqAu} (top and
middle) illustrates the distribution of the spin-mixing parameter
on the Fermi surface of a 10-layer Au(001) film for two
spin-quantization axes, left panel along the [001] direction and right
one along the [100]-direction.  It is easily seen (see middle panel
for finer detail) that spin hot-spots are found near the
edge of the Brillouin zone.  While for most \vc{k}-points inside the
Brillouin zone the spin-mixing parameter has a value of less than 0.05,
at the edge of the Brillouin zone it exhibits very high values
reaching even the maximal value of fully spin-mixed states,
$b^2_{\vc{k}}=\frac{1}{2}$. The anisotropy of the spin-mixing
  parameter is already disclosed by the sheer observation of the
  difference between the color-coded textures of $b^2_{\vc{k}}$ in the
  left panel of Fig.~\ref{anibsqAu} ($\sqa\parallel[001]$) and the
right panel ($\sqa\parallel[100]$), especially if one focuses on the
hot-spot. 
The symmetry of the distribution of $b^2_{\vc{k}}$ depends also on the choice of the spin-quantization axis. 
From Fig.~\ref{anibsqAu} (top) it is obvious that the
  [100]-direction lowers the symmetry with respect to the
  [001]-direction leading to an asymmetric distribution of
  the values of $b^2_{\vc{k}}$.

Summing up over the Fermi surface, we obtain $b^2([001])=0.065$
for Au(001). This value is considerably higher than the value of 0.03
that we find for bulk Au or of 0.035 that we find for a 10-layer
Au(111) film; (111) films show no hot-spots but have Rashba-type
surface states\cite{lashell96,henk03} causing an enhanced spin-mixing
parameter.\cite{heers11} On the contrary, in the (001) thin films,
there are no surface states at the Fermi level, yet the value of $b^2$
is higher.  This demonstrates the importance of the spin-flip
hot-spots in this case.  To estimate the contribution of hot-spots to
the total spin-mixing parameter, we perform the integration in
Eq.~(\ref{bsq}) only for the {\bf k}-points belonging to a small area
around the hot-spots and quantify the area by its contribution to the
density of states at \EF. We find that an area contributing to 4\%
  of $n(\EF)$ contributes by 41\% to the $b^2$ for Cu; an area
  contributing to 9\% of the DOS contributes to 14\% of the $b^2$ for
  Ag; and an area contributing to 10\% of $n(\EF)$ contributes by 52\%
  to the $b^2$ for Au. 

In the case that \sqa\ is along the [100] direction
(Fig.~\ref{anibsqAu}, right-side top and middle panels) the values of
$b_{\mathbf k}^2$ change; the maximum value is then 0.44 and the
integrated one is $b^2([100])=0.042$ for Au.  
Among all directions of \sqa\ in the unit sphere we find that $b^2$ is
maximal for $\sqa\parallel[001]$ and minimal for $\sqa\parallel[100]$
and thus
we obtain for the anisotropy
$\mathcal{A}=\left[b^2([001])-b^2([100])\right]/b^2([100])$ a value of
50\% for Au. This value is gigantic compared to the negligible anisotropy in
the bulk of noble metals, and it is comparable in magnitude to the
anisotropy in e.g.\ W(001) films,\cite{Long13} where it arises from
surface states, or to the anisotropy in  bulk hcp Os
(59\%)\cite{Zimmermann12} where it arises from larger spin-flip hot
areas. However, the present value is still an order of magnitude lower than the one of bulk hcp Hf (830\%) where it
  arises from hot loops at the edge of the hcp Brillouin
  zone, which occur when the Fermi surface crosses
  the hexagonal Brillouin zone edge in hcp metals.\cite{Zimmermann12}

The Fermi surfaces of the 10-layer Cu(001) and Ag(001) films are
  shown in the bottom panel of Fig.~\ref{anibsqAu} together with
  $b^2_{\vc{k}}([001])$ in a color code. Just as in Au, also here the
  spin-flip hot-spots are present at the surface Brillouin-zone edge. The
  hot spots are, however, less intense due to the weaker spin-orbit
  coupling of Cu and Ag.  The anisotropy $\mathcal{A}$ is found to be
  30\% and 8\% for 10-layer Cu(001) and Ag(001) films,
  respectively. The Fermi surface integrated $b^2([001])$ in Cu(001)
  is 0.002, which exceeds the value of 0.0016 that we find for the
  10-layer (111) film in spite of the Rashba surface states of the
  latter, i.e., we see the same qualitative behavior that we observed
  when comparing Au(001) with Au(111); the same holds for Ag.  Our
  results are summarized in Table~\ref{table}. 

We also examine shortly the question of stability of the hot spots
  and of the anisotropy with respect to temperature. We gain a
  qualitative picture by calculating $b^2$ and $\mathcal{A}$ at
  energies slightly different than \EF, in particular $\EF\pm 13.6$ meV
  (13.6 meV correspond to a temperature of 157 Kelvin). Our results are
  summarized in Table~\ref{tableII} and show that there is a
  quantitative change of the values in this energy window, but from
  an order-of-magnitude point of view the results are stable. The
  changes arise because the position and intensity of the hot spots
  varies as the constant-energy surfaces change with energy.

From the results of Tables~\ref{table} and ~\ref{tableII} it seems
  that Ag has an unexpectedly low value of $b^2$ and of $\mathcal{A}$,
  comparable to Cu or even smaller, despite the stronger spin-orbit
  coupling of Ag compared to Cu. We were able to trace this back to the well-known low position of the $d$-bands of Ag with respect to $\EF$, compared
  to Cu or Au, by conducting a numerical experiment.
Acting with a repulsive projection potential, we shifted the Ag $d$-bands higher
  in energy by 1.2~eV, positioning $d$ band edge at 1.5~eV under
  $\EF$, as is approximately the calculated value for Cu and Au. 
The value of $b^2$ increased then by an order of magnitude and the anisotropy increased to 140\%. 
We conclude that the $d$ admixture of the Fermi surface contributes to the value of $b^2$ significantly.

\begin{table}[t]
\begin{tabular}{cllllrr}
\hline
\hline
Metal \ \ &        &          &   $b^2\times 10^2$  &     &
\hspace{12pt}    $\mathcal{A}$ \ \    \\
\hline
      &  bulk      &  \ \ film      & \ \  film  & \ \ film  & \hspace{12pt}           \\
      &\hspace{0pt}        &\hspace{0pt}$\sqa\parallel[001]$ &\hspace{0pt} $\sqa\parallel[100]$ &\hspace{0pt} $\sqa\parallel[110]$ &   \\
\hline
Cu    &\hspace{0pt} 0.15 \hspace{12pt}  &\hspace{0pt} 0.241   &\hspace{0pt} 0.186   &\hspace{0pt} 0.186   & \hspace{0pt} 30\%   \\
Ag    &\hspace{0pt} 0.17 \hspace{12pt}  &\hspace{0pt} 0.193   &\hspace{0pt} 0.179   &\hspace{0pt} 0.179   & \hspace{0pt} 8\%    \\
Au    &\hspace{0pt} 3.0  \hspace{12pt}  &\hspace{0pt} 6.53    &\hspace{0pt} 4.34    &\hspace{0pt} 4.50    & \hspace{0pt} 49\%   \\
\hline
\end{tabular}
\caption{
  Values of the Elliott-Yafet parameter ($\times 10^2$) in bulk and in 10-layer Cu, Ag and Au (001)-films. In the case of the films, the values are given for three directions of \sqa\ with respect to the crystallographic axes ([001] corresponds to the direction of \sqa\ normal to the film surface). The anisotropy values $\mathcal{A}$ correspond to the films. The anisotropy in bulk is less than 0.1\%. \label{table}}
\end{table}

\begin{table}
\begin{tabular}{ccllr}
\hline
\hline
Metal & Energy            & $b^2\times 10^2$ & & $\mathcal{A}$\ \  \\
         &                       & $\sqa\parallel [001]$ & $\sqa\parallel [100]$ & \\
\hline
         & $\EF-13.6$ meV \ \  & \ \  0.275 &\ \  0.200 & 37\% \\
 Cu    & \EF                \ \  & \ \ 0.241 &\ \  0.186 & 29\% \\
         & $\EF+13.6$ meV \ \ & \ \ 0.447 &\ \  0.259 & 72\% \\
\hline
         & $\EF-13.6$ meV \ \  & \ \ 0.249 &\ \  0.200 & 24\% \\
 Ag    & \EF                \ \  & \ \ 0.193 &\ \  0.179 &  8\% \\
         & $\EF+13.6$ meV \ \  & \ \ 0.246 &\ \  0.200 & 23\% \\
\hline
         & $\EF-13.6$ meV \ \  & \ \ 7.36  & \ \ 4.56  & 61\% \\
 Au    & \EF                \ \  & \ \ 6.53  & \ \  4.34  & 50\% \\  
         & $\EF+13.6$ meV \ \  & \ \ 6.01  & \ \  4.12  & 46\% \\
\hline
\end{tabular}
\caption{
Variance of the Elliott-Yafet parameter and its anisotropy with
respect to the energy around \EF\ for Cu, Ag, and Au 10-layer (001)
films. Shown are the values of $10^2\times b^2$ for $\sqa\parallel [001]$ and $[100]$ and the anisotropy at \EF\ and $\EF\pm13.6$
meV. Note that 13.6 meV corresponds to a temperature of 157~K.
\label{tableII}}
\end{table}

Concluding the discussion on the noble-metal films, we comment on the
absence of hot spots in (111) oriented thin films. Part of the reason
is that the (111) oriented atomic planes of fcc crystals are more close-packed than the
(001) oriented atomic planes, resulting in a larger surface
Brillouin-zone area by a factor $2/\sqrt{3}$. 
The fact that the (111) surface Brillouin-zone 
is hexagonal, and in this sense closer in shape to the maximal
circle forming the equator of the bulk Fermi surface, is another aspect. As a result of
both, the projection of the bulk Fermi surface almost fits into the
surface Brillouin-zone, leaving only little room for crossing the zone
boundary. Whether such crossings appear and lead to hot spots has to be
tested for each material and thickness separately, but as we find, it
is not the case for the ultrathin (111) oriented noble metal films.

Finally, for completeness, we briefly discuss the spin-mixing
enhancement in alkali-metal thin films, even though they are typically
not used in spintronics devices.  Although the electronic structure of
alkali metals consists basically of $s$-electrons, even for the alkali
metals the Fermi surface has a non-vanishing $p$ and $d$ character
that is responsible for spin-orbit coupling with strength
$\xi_\ell=\langle
\ell|\frac{\hbar}{2m^2c^2r}\frac{dV}{dr}|\ell\rangle$, with the
  angular momentum state $|\ell\rangle$ being $\ell=1$ or $\ell=2$ for
  $p$ or $d$ wavefunctions, respectively. The Coulomb potential of the
  nucleus that causes most of the spin-orbit coupling is well
screened by the filled shells of the core electrons, contrary to the
noble metals, where a larger part of the screening is done by the
valence electrons and by the not-fully-localized $d$
band. Additionally, the $d$ character of the alkali-metal Fermi
surfaces is less pronounced in comparison to noble metals (with the
exception of Ag) and the spin-orbit coupling of the $d$ states at \EF\
in the noble metals is strong because of the high localization of the
$d$ bands. As a consequence, the spin-orbit strength in alkali metals
is expected to be lower than in the noble metals. Still, we found that
at some film thicknesses, e.g.\ 10 layers of Na(001), K(001), and
Rb(001), the Fermi surface without spin-orbit coupling is degenerate
at the Brillouin-zone edge meaning that the first Fourier component of
the periodic potential vanishes (at least to numerical accuracy, which
we have cross-checked using the full-potential linearized augmented
plane wave method\cite{FLAPW}). In this case spin-orbit coupling
causes a splitting with full spin mixing when \sqa\ is perpendicular
to the film, i.e.\ hot spots with $b_{\mathbf{k}}^2=\frac{1}{2}$
emerge at the Brillouin-zone edge. Yet the magnitude of
$b_{\mathbf{k}}^2$ drops very quickly as the bands separate with
increasing distance from the edge, e.g.\ in Rb(001) we find that
$b_{\mathbf{k}}^2=0.02$ already at a distance of
$0.0005\times\frac{2\pi}{a_{\rm lat}}$ (where $a_{\rm lat}$ is the
lattice constant). Thus in the alkali metal films almost the
  entire magnitude value of $b^2$ comes from a very small region
  around the hot spots (similarly to the case of bulk
  Al\cite{fabian98}) and the anisotropy, that is generally more
  pronounced at and around the hot spots, is significant. In 10-layer
  films, where hot spots are present, we find the following maximal
  and minimal values: $b^2([001])=0.086\times10^{-2}$,
  $b^2([110])=0.025\times10^{-2}$ and $\mathcal{A}=244\%$ for Na(001);
  $b^2([001])=0.11\times10^{-2}$, $b^2([110])=0.039\times10^{-2}$ and
  $\mathcal{A}=182\%$ for K(001); and $b^2([001])=0.45\times10^{-2}$,
  $b^2([110])=0.16\times10^{-2}$ and $\mathcal{A}=181\%$ for Rb(001).

\section{Summary}
In summary, we have shown that the Fermi surface of monovalent metals
in an ultrathin film geometry can show spin-flip hot spots as the
Fermi rings cross the surface Brillouin-zone boundary. This is in
contrast to the bulk of such metals, where it is
known\cite{fabian98,fabian99,fabian_jap99} that hot spots do not
occur, as the Fermi surface is included within the Brillouin zone. We
have furthermore shown that the hot spots contribute to large
anisotropy values of the spin-mixing parameter with respect to the
relative orientation between the spin-quantization axis and the
crystallographic directions. Since the presence of hot spots strongly
influences the spin-relaxation time or the spin-Hall conductivity, our
findings can have consequences in spintronics applications, in
particular since ultrathin noble-metal films are used to transmit or
probe spin currents. The calculated anisotropy can very likely lead to
a variation of the spin-relaxation time with
respect to the spin-polarization direction of the spin current in
experiments and it is important to average this quantity for the estimation of 
those transport properties in polycrystalline samples.
\begin{acknowledgments} 

We acknowledge funding from the project MO 1731/3-1, the programme SPP 1538 SpinCaT of the Deutsche Forschungsgemeinschaft and from the HGF-YIG VH-NG-513 project of the Helmholtz Gemeinschaft. 
We acknowledge the J\"ulich Supercomputing Centre for providing us with computational time.

\end{acknowledgments}

\ \\

\section*{Appendix}

Here we present the algorithm for choosing the linear combination of
degenerate states that maximizes the value of $S_z(\vc{k})$ [see the
discussion after Eq.~(\ref{mixwavefunction})]. Given two
orthogonal solutions at $\vc{k}$, say $\Psi_{\vc{k}}^{(1)}$ and
$\Psi_{\vc{k}}^{(2)}$ (that are found in the process of solving the
eigenvalue problem), any linear combination
$\Psi_{\vc{k}}=c_{\vc{k}}^{(1)}\Psi_{\vc{k}}^{(1)}+c_{\vc{k}}^{(2)}
\Psi_{\vc{k}}^{(2)}$ is still a Bloch eigenfunction of the
Hamiltonian, as long as $c_{\vc{k}}^{(1,2)}$ are complex numbers
independent of $\vc{r}$. Since by normalization
$|c_{\vc{k}}^{(1)}|^2+|c_{\vc{k}}^{(2)}|^2=1$, and since a global
phase factor is irrelevant, we can replace the two complex numbers by
two real parameters $\alpha_{\vc{k}}$ and $\beta_{\vc{k}}$ such that
$\Psi_{\vc{k}}=\cos\frac{\alpha_{\vc{k}}}{2}\Psi_{\vc{k}}^{(1)}+e^{i\beta_{\vc{k}}}\sin\frac{\alpha_{\vc{k}}}{2}
\Psi_{\vc{k}}^{(2)}$. Using a shorthand notation we define the spin
expectation value along $z$ for this linear combination as $S =
\langle\Psi_{\vc{k}}|\frac{\hbar}{2}\sigma_z|\Psi_{\vc{k}}\rangle$ and
analogously $S_{1} =
\langle\Psi_{\vc{k}}^{(1)}|\frac{\hbar}{2}\sigma_z|\Psi_{\vc{k}}^{(1)}\rangle$,
$S_{2} =
\langle\Psi_{\vc{k}}^{(2)}|\frac{\hbar}{2}\sigma_z|\Psi_{\vc{k}}^{(2)}\rangle$
as well as the cross-term $S_{12} =
\langle\Psi_{\vc{k}}^{(1)}|\frac{\hbar}{2}\sigma_z|\Psi_{\vc{k}}^{(2)}\rangle$. Then
we have $S=S_1\cos^2\frac{\alpha_{\vc{k}}}{2} +
S_2\sin^2\frac{\alpha_{\vc{k}}}{2} + (e^{i\beta_{\vc{k}}}S_{12}
+e^{-i\beta_{\vc{k}}}S_{12}^*)
\cos\frac{\alpha_{\vc{k}}}{2}\sin\frac{\alpha_{\vc{k}}}{2}
$. Maximizing or minimizing this expression with respect to
$\alpha_{\vc{k}}$ and $\beta_{\vc{k}}$ gives by definition
$\Psi_{\vc{k}}=\Psi_{\vc{k}}^+$ or $\Psi_{\vc{k}}=\Psi_{\vc{k}}^-$,
i.e., the sought-after states. Demanding that the derivatives with
respect to $\alpha_{\vc{k}}$ and $\beta_{\vc{k}}$ vanish, we arrive at
the result
$\beta_{\vc{k}}=-S_{12}/|S_{12}|+n\pi\equiv-\arg(S_{12})+n\pi$ ($n$
integer), $\alpha_{\vc{k}}=\pm \arctan[2|S_{12}|/(S_1-S_2)]$ (or
$\alpha_{\vc{k}}=\pm \pi$, if $S_1=S_2$), which maximizes or minimizes
the $S_z(\vc{k})$ and which we use for the
$\Psi_{\vc{k}}^{\pm}$. Obviously this has to be repeated for every
$\vc{k}$ on the Fermi surface. The same procedure can be followed for
maximizing the spin along any SQA $\sqa$ by replacing $\sigma_z$ by $\boldsymbol{\sigma}\cdot\sqa$.

\bibliography{apssamp}

\end{document}